\documentclass{elsart4-1}

 \usepackage{graphicx}

\usepackage{bm}
\usepackage{mathrsfs}
\usepackage{amsfonts}

\usepackage{amssymb}

\usepackage[english,francais]{babel}

\newtheorem{e-proposition}[theorem]{Proposition}

\newtheorem{e-definition}[theorem]{Definition\rm}

\renewcommand{\theequation}{\arabic{equation}}
\def\og{\leavevmode\raise.3ex\hbox{$\scriptscriptstyle\langle\!\langle$~}}
\def\fg{\leavevmode\raise.3ex\hbox{~$\!\scriptscriptstyle\,\rangle\!\rangle$}}

\begin{document}

\centerline{Physics or Astrophysics/Header}
\begin{frontmatter}


\selectlanguage{english}
\title{The thermoelectric working fluid: thermodynamics and transport}


\selectlanguage{english}
\author[authorlabel1,authorlabel2]{Giuliano Benenti},
\ead{giuliano.benenti@uninsubria.it}
\author[authorlabel3,authorlabel4,authorlabel5]{Henni Ouerdane}
\ead{henni.ouerdane@rqc.ru}
\author[authorlabel3]{Christophe Goupil}
\ead{christophe.goupil@univ-paris-diderot.fr}

\address[authorlabel1]{Center for Nonlinear and Complex Systems,
Dipartimento di Scienza e Alta Tecnologia,
Universit\`a degli Studi dell'Insubria, via Valleggio 11, 22100 Como, Italy}
\address[authorlabel2]{Istituto Nazionale di Fisica Nucleare, Sezione di Milano,
via Celoria 16, 20133 Milano, Italy}
\address[authorlabel3]{Laboratoire Interdisciplinaire des Energies de Demain (LIED), 
UMR 8236 Universiti\'e Paris Diderot, CNRS, 5 Rue Thomas Mann, 75013 Paris, France}
\address[authorlabel4]{Russian Quantum Center, 100 Novaya Street, Skolkovo, Moscow region 143025, 
Russian Federation}
\address[authorlabel5]{UFR LVE Universit\'e de Caen Normandie, Esplanade de la Paix 14032 Caen, France}


\begin{center}
{\small Received *****; accepted after revision +++++}
\end{center}

\begin{abstract}
Thermoelectric devices are heat engines, which operate as generators or refrigerators using the conduction electrons as a working fluid. The thermoelectric heat-to-work conversion efficiency has always been typically quite low, but much effort continues to be devoted to the design of new materials boasting improved transport properties that would make them of the electron crystal--phonon glass type of systems. On the other hand, there are comparatively few studies where a proper thermodynamic treatment of the electronic working fluid is proposed. The present article aims to contribute to bridge this gap by addressing both the thermodynamic and transport properties of the thermoelectric working fluid covering a variety of models, including interacting systems.

\vskip 0.5\baselineskip

\selectlanguage{francais}
\noindent{\bf R\'esum\'e}
\vskip 0.5\baselineskip
\noindent
{\bf Le fluide de travail thermo\'electrique: thermodynamique et transport. }
Les dispositifs thermo\'electriques sont des machines thermiques pouvant op\'erer en mode g\'en\'erateur ou r\'efrig\'erateur en utilisant les \'electrons de conduction comme fluide de travail. Le rendement de conversion chaleur-travail a toujours \'et\'e typiquement bas, mais la conception de nouveaux mat\'eriaux thermo\'electriques est l'objet d'efforts cons\'equents en vue d'obtenir des syst\`emes de type crystal \'electronique-verre de phonons. Par comparaison, il y a cependant un d\'eficit de traitement approfondi des propri\'et\'es thermodynamiques du fluide de travail thermo\'electrique.  Le pr\'esent article vise \`a contribuer \`a combler cet \'ecart en examinant les propri\'et\'es thermodynamiques et de transport du fluide de travail thermo\'electrique dans le cadre de diff\'erents mod\`eles, incluant les syst\`emes en interaction.

 \vskip 0.5\baselineskip
\noindent{\small{\it Mots-cl\'es~:} Thermo\'electricit\'e; Transitions de phase; Syst\`emes en interaction}}
\end{abstract}
\end{frontmatter}


\selectlanguage{english}
\section{Introduction}
\label{sec:intro}

In a general manner, transport phenomena are irreversible processes: the generation of fluxes within the system upon which external constraints are applied, are accompanied by energy dissipation and entropy production \cite{Pottier2010}. Now, assume a thermodynamic system in which electric transport and heat transport may take place. Thermoelectric effects may be thus viewed as the result of the mutual interaction of two irreversible processes, electrical transport and heat transport, as they take place \cite{Onsager1931,Callen1948}. This mutual interaction may be quantified by the so-called degree of coupling \cite{Kedem1965} upon which the thermoelectric conversion efficiency depends. Owing to the coupling of electrical charges and heat transport (the strength of which is given by the Seebeck coefficient) thermoelectric systems thus form a most interesting class of heat engines, not only for practical purposes, but also from a fundamental viewpoint: Some of the theoretical developments of the mathematical relationships between forces and fluxes in coupled transport rest on the analysis of thermoelectric systems \cite{Onsager1931,Thomson1853} and thermoelectricity is a touchstone for theories of irreversible thermodynamics \cite{Groot1958}. On the practical side, thermoelectric devices may be used to collect and transform waste heat into electrical power, to pump heat for cooling or heating, and for temperature measurement. These devices are particularly reliable, even in hostile places, as demonstrated in the context of deep-space probe missions. Indeed, since thermoelectric phenomena are electronic in nature, their energy conversion efficiency is not system-size-dependent, their operation does not rely on moving parts, and there is no need of refrigerant fluids for coolers. 

The thermoelectric performance is governed by the thermoelectric figure of merit \cite{ioffe,goldsmid}

\begin{equation}
ZT=\frac{\sigma S^2}{\kappa_e+\kappa_{ph}},
\label{eq:ZT-def}
\end{equation}

\noindent where $\sigma$ is the electrical conductivity, $S$ the thermopower (or Seebeck coefficient), $\kappa_e$ and $\kappa_{ph}$ the electronic and the phononic heat conductivities, respectively, and $T$ the system's temperature. To envisage applications for thermoelectric systems other than those for which sustainability and reliability are more important than low-level efficiency and high cost, values of $ZT$ greater than 4 are mandatory \cite{Vining2009}. The great challenge to increase thermoelectric efficiency relies on understanding the microscopic mechanisms that may allow to control individually $\sigma$, $S$, and $\kappa=\kappa_e+\kappa_{ph}$. However, the different transport coefficients are interdependent making optimisation extremely difficult and so far, no clear paths exist which may lead to reach that target. A significant example of this interdependence is the Wiedemann-Franz law \cite{ashcroftmermin} which states that for metallic materials, $\sigma$ and $\kappa_e$ are proportional, thus making metals poor thermoelectric materials in general.

The energy transferred by phonons represents a useless heat leak and recent efforts in materials science and engineering focused on strategies to lower lattice heat conduction \cite{Snyder,Shakouri}, in particular in low-dimensional nanostructures where rough surfaces can efficiently scatter phonons \cite{Majumdar,Bourgeois}. On the other hand, even ideally reducing $\kappa_{ph}\to 0$, as in the electron crystal-phonon glass paradigm \cite{Slack}, would not by itself guarantee $ZT\gg 1$, due to the remaining electronic contribution $\kappa_e$ to the thermal conductivity. However it must be clearly mentionned that $\kappa_{e}$ accounts for two physical phenomena: heat transfer by conduction (Fourier's law) and heat transfer by electron \emph{convection} \cite{Apertet2012Conv}, which is the so-called Peltier term of the heat flux \cite{Goupil2011}. Note that this latter represents the actual and global electronic movement within the conduction submitted to a thermal gradient, hence convection only can be seen as the ``useful'' contribution to heat transfer across the system. Here, we will thus rather focus on an aspect of thermoelectricity, which is nearly always neglected: investigate suitable strategies to improve the properties of the thermoelectric working fluid itself. 

The article is organized as follows: in Section 2, we cover the thermodynamics and transport properties of the noninteracting working fluid. We then focus on phase transitions in Section 3 and finally consider interacting working fluids in Section 4, covering momentum conserving systems. In Section 5, we discuss the question of optimization of device operation, as the system's working fluid is dissipatively coupled to a hot and to a cold temperature bath, and we end the article with concluding remarks.

\section{The noninteracting thermoelectric working fluid} 

\subsection{The Onsager approach to coupled transport}

In the linear response regime, the relationship between currents and generalized forces is linear \cite{callen,mazur} and in the case of thermoelectric transport we have 

\begin{eqnarray}
\left\{
\begin{array}{l}
j_e=L_{ee} \mathcal{F}_e + L_{eh} \mathcal{F}_h,
\\
\\
j_h=L_{he} \mathcal{F}_e + L_{hh} \mathcal{F}_h,
\end{array}
\right.
\label{eq:coupledlinear}
\end{eqnarray}

\noindent where $j_e$ is the electric current density, $j_h$ is the heat current density, and the conjugated local generalized forces are given by $\mathcal{F}_e=-\nabla \mu/eT$ and $\mathcal{F}_h=\nabla(1/T)$, $\mu$ is the electrochemical potential and $e<0$ is the electron charge. The coefficients $L_{ab}$ ($a,b=e,h$) are known as kinetic coefficients or Onsager coefficients;
we will denote ${\bm L}$ the Onsager matrix with matrix elements $L_{ab}$. 

The Onsager coefficients are subject to two fundamental constraints. First, the second law of thermodynamics requires the positivity of the entropy production rate, 

\begin{equation}
\dot{s}=\mathcal{F}_e J_e + \mathcal{F}_h J_h=
L_{ee} \mathcal{F}_e^2 + L_{hh} \mathcal{F}_h^2 +
(L_{eh}+L_{he}) \mathcal{F}_e \mathcal{F}_h\ge 0,
\label{eq:sprod}
\end{equation}

\noindent where $s$ is the local entropy density. Eq.~(\ref{eq:sprod}) is equivalent to the conditions 

\begin{equation}
L_{ee}\ge 0,
\quad
L_{hh}\ge 0,
\quad
L_{ee}L_{hh}
-\frac{1}{4}\,(L_{eh}+L_{he})^2 \ge 0.
\label{dots}
\end{equation}

\noindent Second, assuming the property of time-reversal invariance of the equations of motion, Onsager derived \cite{Onsager1931} fundamental relations, known as Onsager reciprocal relations for the cross coefficients of the Onsager matrix: $L_{ab}=L_{ba}$.

The kinetic coefficients $L_{ab}$ are related to the thermoelectric transport coefficients: the electrical conductivity $\sigma$, the thermal conductivity $\kappa$, the Seebeck coefficient $S$, and the Peltier coefficient $\Pi$, as 

\begin{equation}
\sigma=-e\,\left(\frac{j_e}{\nabla\mu}\right)_{\nabla T=0}=\frac{L_{ee}}{T},
\label{eq:el_conductivity}
\end{equation}
\begin{equation}
\kappa=-\left(\frac{j_h}{\nabla T}\right)_{j_e=0}=
\frac{1}{T^2}\frac{\det {\bm L}}{L_{ee}},
\label{eq:th_conductivity}
\end{equation}
\begin{equation}
S=-\frac{1}{e}\left(\frac{\nabla \mu}{\nabla T}\right)_{j_e=0}=
\frac{1}{T}\frac{L_{eh}}{L_{ee}},
\label{eq:seebeck}
\end{equation}
\begin{equation}
\Pi=\left(\frac{j_h}{j_e}\right)_{\nabla T=0}
=\frac{L_{he}}{L_{ee}}.
\label{eq:peltier}
\end{equation}

\noindent For systems with time reversal symmetries $\Pi=TS$ due to the Onsager reciprocal relations \cite{Goupil2011}.

Using  Eqs.~(\ref{eq:ZT-def}) and (\ref{eq:el_conductivity})-(\ref{eq:seebeck}), the thermoelectric figure of merit reads 

\begin{equation} 
ZT=\frac{{L}_{eh}^2}{\det {{\bm L}}}.
\label{ZTOnsager}
\end{equation}

\noindent Thermodynamics only imposes a lower bound on the figure of merit: $ZT\ge 0$, and the thermoelectric conversion efficiency is a monotonous increasing function of $ZT$, the ideal Carnot efficiency being achieved in the limit $ZT\to\infty$. To understand why thermoelectric systems boast rather poor peformance levels in terms of conversion efficiency, and how one may obtain high values of $ZT$, which ideally should be ``only'' greater than 4 for practical purposes, we must compute and analyze the transport coefficients. 

\subsection{Transport parameters} 

For non-interacting systems, Mahan and Sofo showed long ago \cite{mahansofo} that the best thermoelectric efficiency can be obtained in systems with \emph{energy filtering}, namely where the
energy width of the main conducting channel is very narrow. This result can be understood as follows. We can write the Onsager coefficients as \cite{mahansofo,BCSWreview} 

\begin{equation} 
L_{ee}=2 e^2 T K_0,\;\;
L_{eh}=L_{he}=2 e T K_1,\;\;
L_{hh}=2 T K_2,
\end{equation}

\noindent where the factor 2 is due to spin degeneracy, and the integrals 

\begin{equation}
K_n\equiv \int_{-\infty}^\infty
dE (E-\mu)^n \Sigma(E) \left(-\frac{\partial f}{\partial E}\right)
\label{eq:Kn}
\end{equation}

\noindent are written in terms of the Fermi distribution function $f(E)=\{\exp[(E-\mu)/k_B T]+1\}^{-1}$ and of the transport distribution function $\Sigma(E)$, which can be derived both in the semiclassical Boltzmann approach \cite{mahansofo} or using Green's-function techniques\footnote{In the Landauer approach, where conductances rather than conductivities are used, similar expressions are derived with the transport distribution function substituted by the transmission function \cite{BCSWreview,datta,imry}.} \cite{mahanPRB}. We can therefore write the transport coefficients as follows: 

\begin{equation}
\sigma = 2e^2 K_0,\;
\kappa=\frac{2}{T}\left(K_2-\frac{K_1^2}{K_0}\right),\;
S=\frac{1}{eT}\frac{K_1}{K_0}.
\label{eq:boltzmann.ex}
\end{equation} 

\noindent The Seebeck coefficient may be seen as the average value of the entropy involved in the thermoelectric transport, $(E-\mu)/T$, over a probability density function given by the product of the transport distribution function and the energy derivative of $f$. In metals and degenerate semiconductors, where electrons above the Fermi level carry a heat current that is practically the opposite of that carried by the electrons below the Fermi level, the Seebeck coefficient is typically small: since $E-\mu$ changes sign as $E$ varies, it is essential that $\Sigma$ presents an asymmetric profile~\cite{Adel}, to avoid cancellation of $K_1$.

The Carnot efficiency is achieved in the case of energy filtering, i.e. when the transmission is possible only within a tiny energy window around a value $E=E_\star$. Indeed, in this case from Eq.(\ref{eq:Kn}) we obtain $K_n\approx (E_\star-\mu)^n K_0$, and therefore 

\begin{equation}
ZT=\frac{\sigma S^2}{\kappa}\,T=\frac{K_1^2}{K_0 K_2-K_1^2}\to\infty.
\end{equation}

On the other hand, the Wiedemann-Franz law is recovered in the limit of a broad and smooth transmission function. More precisely, we consider the Sommerfeld expansion \cite{ashcroftmermin} of integrals (\ref{eq:Kn}) to the leading order in $k_B T/E_F$, with $k_B$ being the Boltzmann constant and $E_F=\mu(T=0)$ being the Fermi energy. Such an expansion is valid for a smooth function 
$\Sigma(E)$. The transport distribution function is approximated as follows: 

\begin{equation}
\Sigma (E) \approx \Sigma (\mu) + \left.\frac{d \Sigma (E)}{dE}\right|_{E=\mu}
(E-\mu ).
\end{equation}

\noindent After inserting this expansion into (\ref{eq:Kn}), we obtain the leading order terms of the Sommerfeld expansion of integrals $K_n$: 

\begin{equation}
K_0\approx {\Sigma(\mu)},\quad
K_1\approx \frac{\pi^2}{3}\,(k_B T)^2
\left.\frac{d \Sigma (E)}{dE}\right|_{E=\mu},\quad
K_2\approx  \frac{\pi^2}{3}\,(k_B T)^2 \Sigma(\mu).
\label{eq:K0K1K2}
\end{equation}

In this derivation, we have used the fact that $\partial f/\partial E$ is an even function of $\epsilon\equiv (E-\mu)/k_B T$. Hence, $K_0$ and $K_2$ are determined to the leading order by $\Sigma(\mu)$. In contrast, $(E-\mu)\partial f/\partial E$ is an odd function of $\epsilon$, so that $K_1$ is determined by the derivative $\left.\frac{d \Sigma (E)}{dE}\right|_{E=\mu}$.
We then derive from  Eq.~(\ref{eq:boltzmann.ex})

\begin{equation}
\sigma\approx {2 e^2} \Sigma(\mu),
\quad
\kappa \approx \frac{2 \pi^2 k_B^2 T}{3}\,\Sigma(\mu),
\end{equation}
and from these relations we find the Wiedemann-Franz law
\begin{equation}
\frac{\kappa}{\sigma T} \approx \mathfrak{L} \, ,  \label{wflaw}
\end{equation}
where the constant value
\begin{equation}
\mathfrak{L} = \frac{\pi^2}{3} \left( \frac{k_B}{e} \right)^2  \label{lorenz}
\end{equation}

\noindent is known as the Lorenz number. 

To derive the Wiedemann-Franz law we have considered only the leading order term in the Sommerfeld expansion, i.e. we have neglected in the heat conductivity $K_1^2/K_0$ with respect to $K_2$.
This in turn implies that $L_{ee}L_{hh}\gg L_{eh}^2$ and the thermal conductivity $\kappa\approx L_{hh}/T^2$. When the Wiedemann-Franz law is valid, it is not possible to obtain large
thermoelectric efficiency, as in this case the figure of merit $ZT=L_{eh}^2/{\det{\bm L}}\approx L_{eh}^2/L_{ee}L_{hh}\ll 1$. Consequently, to get large values of $ZT$ one should search for physical situations where the Wiedemann-Franz law is violated. For non-interacting particles, violations can occur in small systems where transmission shows a significant energy dependence
\cite{stone,shakouri2007,bss10,bjs10,bbb12,sanchezprb13} or in bulk systems in the vicinity of a phase transition, where the transport distribution function is not an analytic function.

\subsection{Anderson transition}
\label{Anderson}

We consider first the metal-insulator transition in the three-dimensional Anderson model, and we define $\sigma_0(E)=2 e^2 \Sigma(E)$, so that 

\begin{equation}
\sigma=\int_{-\infty}^\infty
dE  \sigma_0(E) \left(-\frac{\partial f}{\partial E}\right),
\label{eq:sigma0}
\end{equation}

\noindent In this case, $\sigma_0(E)$ corresponds to the $T=0$ electrical conductivity of the system when the Fermi energy $E_F=E$. In the Anderson transition, a mobility edge $E_m$ separates localized states (for $E<E_m$) from extended states (for $E>E_m$), and the zero temperature conductivity changes non-analytically at $E=E_m$: 

\begin{equation}
\sigma_0(E)=
\left\{
\begin{array}{l}
A(E-E_m)^x,\;\;\hbox{if }E\ge E_m,
\\
\\
0,\;\;\hbox{if }E\le E_m,
\end{array}
\right.
\end{equation}

\noindent where $A$ is a constant and $x$ the conductivity critical exponent whose value is unknown, in spite of several analytical and numerical methods used to attemp its evaluation \cite{imryanderson}.

We can here express the thermopower $S$, the ratio $\kappa/\sigma T$ and the figure of merit $ZT$ in terms of a single scaling parameter, 

\begin{equation}
z\equiv \frac{\mu-E_m}{k_B T}.
\end{equation}

We then calculate: 

\begin{equation}
S=\frac{k_B}{e}~
\frac{\displaystyle \int_{-z}^\infty dy y (y+z)^x f^\prime (y)}
{\displaystyle \int_{-z}^\infty dy (y+z)^x f^\prime (y)},
\end{equation}

\noindent with $f'(y)=-1/[4\cosh^2(y/2)]$ derivative of the Fermi function, and 

\begin{equation}
\frac{\kappa}{\sigma T}=\left(\frac{k_B}{e}\right)^2~
\frac{\displaystyle \int_{-z}^\infty dy y^2 (y+z)^x f^\prime (y)}
{\displaystyle \int_{-z}^\infty dy (y+z)^x f^\prime (y)}-S^2,
\end{equation}

\noindent so that also the figure of merit $ZT$ only depends on the scaling parameter $z$. 

\begin{figure}
\center
\centerline{\includegraphics[scale=0.4]{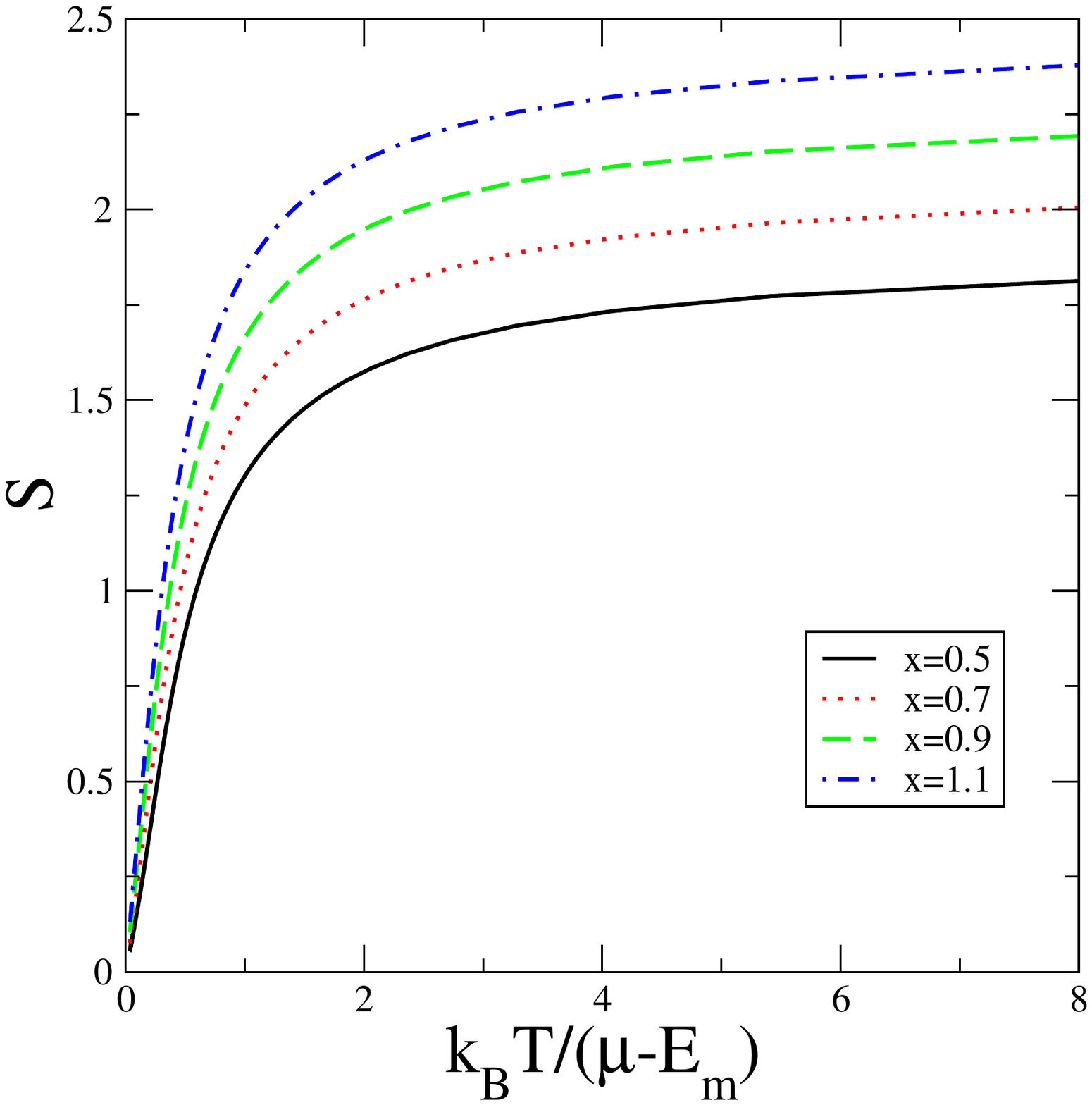}\includegraphics[scale=0.4]{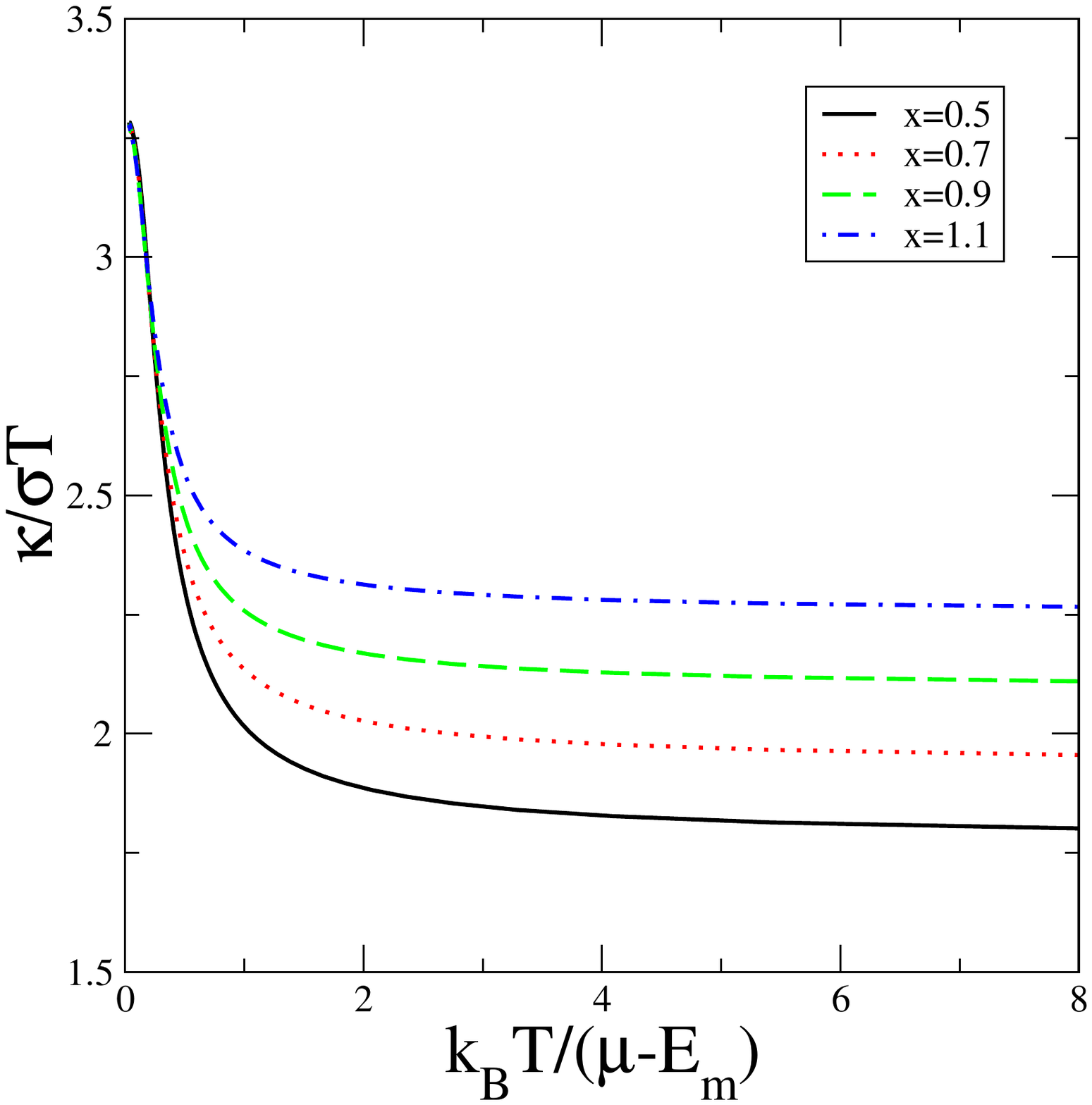}}
\vglue -4cm 
\centerline{\includegraphics[scale=0.4]{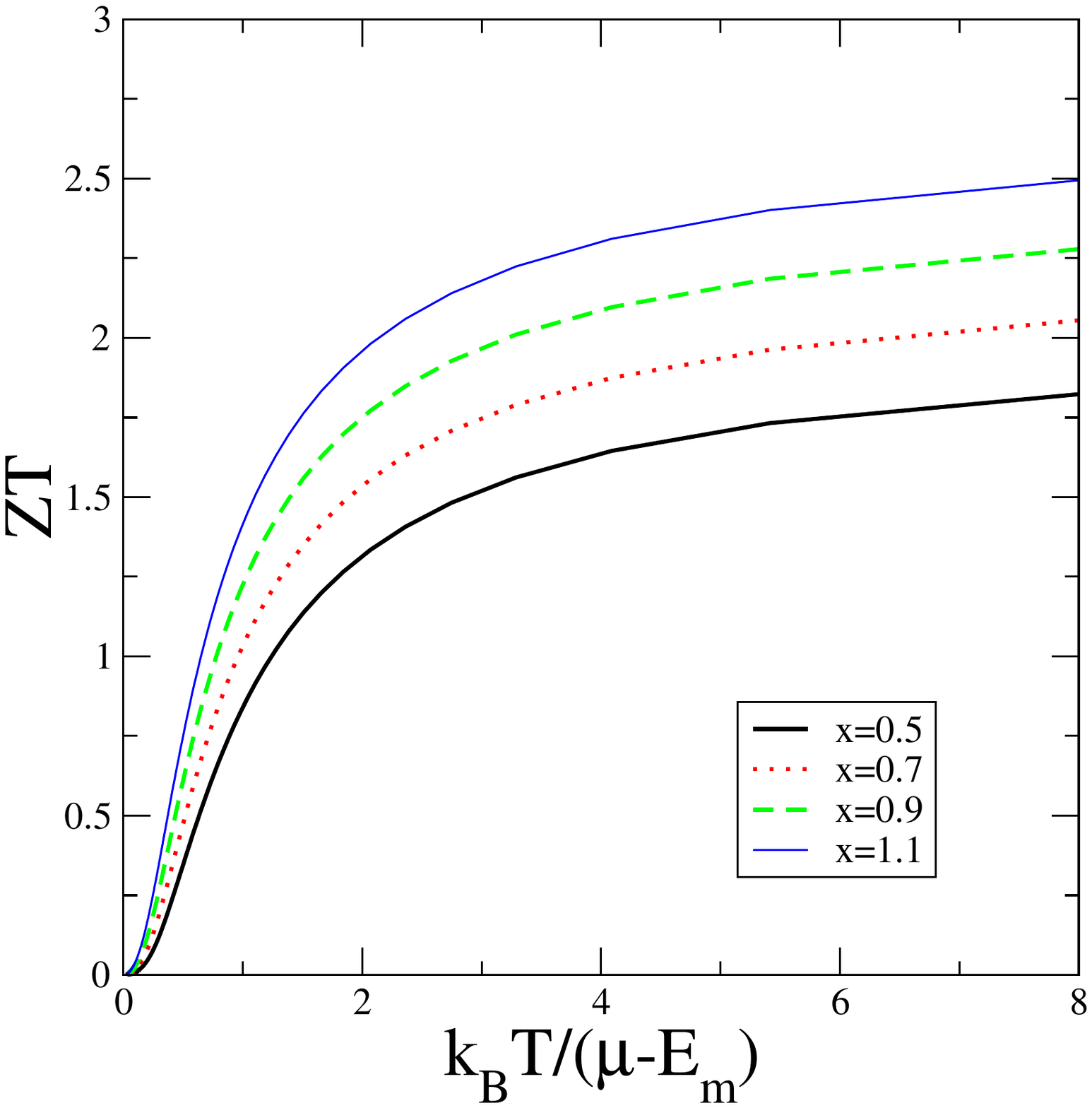}}
\caption{Thermopower (in units of $k_B/e$) (top left), 
ratio $\kappa/\sigma T$ (in units of $(k_B/e)^2$) (top right),
and thermoelectric figure of merit $ZT$ (bottom)
as a function of the inverse of the scaling parameter $z=(\mu-E_m)/k_BT$,
for different values of the critical exponent $x$.}
\label{fig:Anderson}
\end{figure}

The ratio $\kappa/\sigma T$, the figure of merit $ZT$ and the thermopower\footnote{See Ref.~\cite{imryanderson} also for a detailed discussion of thermopower close to the Anderson transition.} $S$, are depicted in Fig.~\ref{fig:Anderson}, for various values of the critical exponent $x$. For low enough temperatures, that is, for $k_BT\ll \mu-E_m$, the nonanalytic behavior of $\sigma_0(E)$ at the mobility edge is not relevant. We can therefore apply the Sommerfeld expansion and obtain 

\begin{equation}
S\approx \frac{\pi^2}{3} \frac{k_B}{e}
\left(\frac{x}{z}\right),\quad 
\frac{\kappa}{\sigma T}\approx \mathfrak{L}, \quad
ZT\approx \frac{\pi^2}{3}\left(\frac{x}{z}\right)^2.
\end{equation}

\noindent On the other hand, increasing the temperature the non-analyticity at $E_m$ plays an important role, and the Sommerfeld expansion can no longer be applied; the 
Wiedemann-Franz law is then violated and relatively large values of $ZT$ may be obtained. More interesting enhancements of $ZT$ can be obtained close to electronic phase transitions in interacting systems.

\section{Phase transitions}
\label{sec:phaseint} 

A reasoning by Vining \cite{Vining1997} suggests that large values of $ZT$ can be expected near electronic phase transitions. Fist of all, we consider the thermal conductivity $\kappa'$ at zero voltage ($\mathcal{F}_e=0$), which is related to the thermal conductivity $\kappa$ measured at zero electric current as $\kappa=\kappa'+\sigma S\Pi$. The thermoelectric figure of merit can then be written as 

\begin{equation}
ZT=\gamma_K-1,
\quad \gamma_k\equiv \frac{\kappa'}{\kappa};
\end{equation}

\noindent and we see that $ZT$ diverges if and only if the ratio $\gamma_k$ diverges. Of course, there is no such a thing as a liquid-gas phase transition in electronic systems, but other types of transitions, involving electron pairing mechanism as is the case for superconductivity, exist, and it is worthwhile to focus now on the thermodynamic properties of the working fluid itself rather than on transport. 

Vining's initial idea \cite{Vining1997} was refined and developed by Ouerdane \emph{et al} \cite{Ouerdane2015}: the rationale of their analysis rests on the facts that thermoelectric transport is essentially a convective process \cite{Apertet2012Conv}, and that convection may be enhanced in the vicinity of a phase transition. The key point was then to see how to characterize the thermoelastic properties of the electronic working fluid and its capacity for enhanced convective transport. This they did by introducing the \emph{thermodynamic figure of merit}, the meaning of which is physically transparent: it is related to the compressibility of the considered fluid and the Prandtl number, given by the ratio of the kinematic viscosity over the thermal diffusivity \cite{Blundell}. As a matter of fact, this characteristic number in fluid dynamics provides a direct link between the thermodynamic properties of the fluid and its capacity for convective transport. 

Now, consider an open system characterized by the number $N$ of particles, the electrochemical potential $\mu$ and the temperature $T$. We have \cite{Vining1997}:

\begin{eqnarray}
\left\{
\begin{array}{l}
dN= \left( \frac{\partial N}{\partial \mu}\right)_T\, d\mu
+ \left( \frac{\partial N}{\partial T}\right)_\mu\, dT \equiv
C_{NN} d\mu + C_{N\mathscr{S}} dT,
\\
\\
d\mathscr{S}=
\left(\frac{\partial \mathscr{S}}{\partial \mu}\right)_T\,d\mu 
+ \left(\frac{\partial \mathscr{S}}{\partial T}\right)_\mu\,dT \equiv 
C_{\mathscr{S}N} d\mu + C_{\mathscr{S}\mathscr{S}} dT,
\end{array}
\right.
\label{eq:capacitymatrix}
\end{eqnarray}

\noindent where $\mathscr{S}$ denotes the system's entropy. These equations are formally similar the the coupled transport equations (\ref{eq:coupledlinear}), with the \emph{capacity matrix} ${\bm C}$ (with matrix elements $C_{ab}$, $a,b=N,\mathscr{S}$) rather than the Onsager matrix ${\bm L}$. Note that $C_{N\mathscr{S}}=C_{\mathscr{S}N}$ due to the extended Maxwell relation

\begin{equation}
\left(\frac{\partial N}{\partial T}\right)_\mu=
\left(\frac{\partial \mathscr{S}}{\partial \mu}\right)_T.
\end{equation}

\noindent Moreover, $C_{\mathscr{S}\mathscr{S}}\equiv C_\mu$ is the entropy capacity at constant $\mu$. Finally, the entropy capacity at constant $N$ is

\begin{equation}
C_{N}\equiv\left(\frac{\partial \mathscr{S}}{\partial T}\right)_N
=\frac{\det{\bm C}}{C_{NN}},
\end{equation}

\noindent where the last equality is derived after setting $dN=0$ in (\ref{eq:capacitymatrix}).

We now consider a thermodynamic cycle consisting of two constant electrochemical potential strokes $d\mu$ apart and two constant particle number strokes $dN$ apart. The infinitesimal work performed by this cyclic process is $-d\mu dN$ and we can compare it with the work $d\mathscr{S} dT$ performed by a Carnot cycle consisting of two isothermal strokes $dT$ apart and two adiabatic
strokes $d\mathscr{S}$ apart. The ratio between the heat to work conversion efficiencies of the above two processes is therefore given by 

\begin{equation}
\frac{\eta}{\eta_C}=\frac{-d\mu dN}{d\mathscr{S} dT}.
\end{equation} 

\noindent As the Carnot efficiency for a cycle operating between temperatures $T$ and $T+dT$ is $\eta_C=dT/T$, we obtain 

\begin{equation}
\eta=\frac{-d\mu dN}{Td\mathscr{S}}
=\frac{-d\mu (C_{NN} d\mu + C_{N\mathscr{S}} dT)}{T
(C_{\mathscr{S}N} d\mu + C_{\mathscr{S}\mathscr{S}} dT)}.
\end{equation}

\noindent and, similarly to thermoelectric transport, one can show \cite{Ouerdane2015} that the maximum efficiency of this thermodynamic cycle is a monotonous growing function of the thermodynamic figure of merit: 

\begin{equation}
Z_{\rm th} T=\frac{C_{N\mathscr{S}}^2}{\det {\bm C}}= 
\gamma_{\mu N}-1, \quad \gamma_{\mu N}\equiv \frac{C_\mu}{C_N}.
\end{equation}

\noindent We point out that $Z_{\rm th} T$ is purely determined by the properties of the working fluid, without referring to thermoelectric transport. Consequently, it does not include any contribution from phonons, that instead affect the thermoelectric figure of merit $ZT$.

As a final step, we use the mapping $\mu\rightarrow -p$ and $N\to V$, with $p$ and $V$ pressure and volume of a gas. We then consider the infinitesimal work $dpdV$ performed by a cycle consisting of two isobaric strokes $dP$ apart and two isochoric strokes $dV$ apart and compare it again with the work $d\mathscr{S} dT$ performed by a Carnot cycle. By using the same steps as above for the $\mu-N$ system, we find that the heat to work conversion efficiency is a monotonous function of the thermodynamic figure of merit for the $p-V$ systems: 

\begin{equation}
Z_{\rm th} T=
\gamma_{pV}-1, \quad \gamma_{pV}\equiv \frac{C_p}{C_V},
\end{equation}
where
\begin{equation}
C_p\equiv T \left(\frac{\partial \mathscr{S}}{\partial T}\right)_p,
\quad
C_V\equiv T \left(\frac{\partial \mathscr{S}}{\partial T}\right)_V
\end{equation} 

\noindent are the heat capacity at constant pressure and volume, respectively. For a classical ideal (noninteracting) gas, $1<\gamma_{pV}\le \frac{5}{3}$, with the upper bound achieved for monoatomic gases. Hence, $Z_{\rm th} T \le \frac{2}{3}$. On the other hand, the ratio $\gamma_{pV}$ (and $Z_{\rm th}$) can diverge for condensable gases, at the critical temperature $T_c$ between the gas phase and the two-phase region (gas-liquid coexistence). The analogy with a classical gas suggests the possibility of large values of $Z_{\rm th}$ close to electronic phase transitions, strongly improving the thermoelectric properties of the working fluid with respect to noninteracting systems in their normal state. Indeed, it has been recently demonstrated \cite{Ouerdane2015} that $Z_{\rm th} T$ diverges when approaching from the normal phase the critical point for the transition to the superconducting phase, in the fluctuation regime \cite{Varlamov}. 

More specifically, to analyze the thermodynamic properties of the electronic working fluid near the superconducting phase transition, one needs a set of four thermoelastic coefficients: $\beta N = \left(\partial N/\partial T\right)_{\mu}$: analogue to thermal dilatation coefficient; $\chi_T N= \left(\partial N/\partial \mu\right)_{T}$: analogue to isothermal compressibility; $c_{\mu} N= T\left(\partial \mathscr{S}/\partial T\right)_{\mu}$: analogue to specific heat at constant pressure; $c_N N= T\left(\partial \mathscr{S}/\partial T\right)_{N}$: analogue to specific heat at constant volume. Application of the extended Maxwell's relations yields $\beta/\chi_T = \mathscr{S}_N$ with $\mathscr{S}_N = \left(\partial \mathscr{S}/\partial N\right)_T$, which reflects the notion of entropy per particle introduced by Callen \cite{Callen1948} and the ensuing thermodynamic definition of the thermoelectric coupling: $s_{\rm th}=\beta\chi_T^{-1}/e$. 

The heat capacity at constant particle number can be derived from the knowledge of the free energy of the fluctuation Cooper pairs: $c_N = (-T/N_{\rm cp}) \partial^2\mathcal{F}_{\rm cp}/\partial T^2$, where $N_{\rm cp}$ is the number of fluctuation Cooper pairs; the other three thermoelastic coefficients $\beta$, $\chi_T$, and $c_{\mu}$ are given below. For clarity we give the derivation steps for $\chi_T$; the other two follow similar steps. For a many-particle system with energy distribution function $f$ and density of state $g$, the number of particles $N$ is given by:

\begin{equation}
N = \int_0^{\infty} g(E)f(E){\rm d}E
\end{equation}

\noindent By definition, $\chi_T N= \left(\partial N/\partial \mu\right)_{T}$, and the first partial derivative of $N$ with respect to $\mu$ at constant temperature takes the form:

\begin{equation}
\frac{\partial N}{\partial \mu} = \int_0^{\infty} g(E)\frac{\partial f}{\partial \mu}{\rm d}E = \int_0^{\infty} g(E)\left(-\frac{\partial f}{\partial E}\right){\rm d}E
\end{equation} 

\noindent since the density of state does not depend on $\mu$ and $\partial f/\partial \mu = -\partial f/\partial E$. We thus obtain $\chi_T$ as well as the other two coefficients in the same fashion:

\begin{eqnarray}
\nonumber
\chi_T N&=& \int_0^{\infty} g(E)\left(-\frac{\partial f}{\partial E}\right) {\rm d}E,\\
\nonumber
\beta N&=& \frac{1}{T_{\rm a}}\int_0^{\infty} g(E)\left(E-\mu_{\rm a}\right)\left(-\frac{\partial f}{\partial E}\right) {\rm d}E,\\
\nonumber
c_{\mu} N&=& \frac{1}{T_{\rm a}}\int_0^{\infty} g(E)\left(E-\mu_{\rm a}\right)^2\left(-\frac{\partial f}{\partial E}\right) {\rm d}E,
\end{eqnarray} 

\noindent where $\mu_{\rm a}$ and $T_{\rm a}$ are the average values of the electrochemical potential and temperature accross the considered system. The shape of the derivative of the Fermi function is given in Section~\ref{Anderson}.

As shown in Ref.~\cite{Ouerdane2015}, $Z_{\rm th}$ may diverge at \emph{finite} temperature in the fluctuation regime case, while it does not for the standard Bose and Fermi gases. This explains why thermoelectric devices, which use the noninteracting electron gas a working fluid, are not very efficient energy conversion devices despite the intense efforts to improve their performance over decades. The fluctuation regime studied in Ref.~\cite{Ouerdane2015}, where phonons are put to work to bind electrons, shows that it is possible to prepare highly compressible electrically charged working fluids; but, other electronic systems could also present enhanced thermoelectric properties as long as they boast a high-compressibility factor.

\section{Interacting working fluids}
\label{sec:cmotion}

\subsection{General considerations}

The thermoelectric properties of strongly interacting systems are of great interest since their efficiency is not bounded by limitations due to the Wiedemann-Franz law, which applies for 
bulk non-interacting metallic-like systems. Moreover, experimental results on some strongly correlated materials such as sodium cobalt oxides revealed unusually large thermopower values \cite{Terasaki1997,Wang2003}, in part attributed to strong electron-electron interactions \cite{Peterson2007} due to the $d$ or $f$ character of the band structure in the vicinity of the Fermi level, and also to the fact that oxides have both spin and orbital degrees of freedom, hence high entropy. Further, strong correlations may be acted upon to increase the power factor $S\sigma^2$ through the tuning of crystal-field and spin-orbit coupling up to an optimum as shown for correlated Kondo insulators in Ref.~\cite{Philipps2013}. 

As a matter of fact, fairly little is known about the thermoelectric properties of strongly correlated systems. Theoretical formalisms include the dynamical mean field theory \cite{dmft} for non-perturbative computation of the self-energy of the many-body systems and of the thermoelectric and thermodynamic properties of, e.g. a hole-doped Mott insulator \cite{Deng2013}; the slave boson formalism \cite{Lee1987,Ouerdane2007} to treat the large on-site repulsion term in lattice models as in Ref.~\cite{Philipps2013}; and the Green-Kubo approach employed below. Analytical results are thus rare and numerical simulations (Monte Carlo and numerical renormalization group), even if based on simple effective Hamiltonian models, such as the single-band Hubbard model or the single impurity Anderson model, are challenging. However, on the basis of the Green-Kubo formula, we can discuss a thermodynamic argument suggesting that the Carnot efficiency is achieved in the thermodynamic limit for non-integrable momentum-conserving systems, which we cover hereafter.

\subsection{Momentum-conserving systems}

A general thermodynamic argument \cite{Benenti2013} corroborated by numerical simulations \cite{Benenti2013,Benenti2014,Chen2015} predicts that nonintegrable systems with momentum conservation achieve the Carnot efficiency at the thermodynamic limit. Such an argument is rooted in the Green-Kubo formula, which expresses the Onsager kinetic coefficients in terms of dynamic correlation functions of the corresponding current operators, calculated at thermodynamic equilibrium~\cite{kubo,mahan}: 

\begin{equation}
L_{ab} = \lim_{\omega\to 0} {\rm Re} [L_{ab} (\omega)], 
\quad
L_{ab}(\omega) = \lim_{\epsilon\to 0}
\int_0^\infty dt e^{-i(\omega\!-\!i\epsilon)t}\lim_{\Omega\to\infty}\frac{1}{\Omega}
\int_0^\beta d\tau\langle \hat{J}_a \hat{J}_b (t+i\tau)\rangle, 
\label{eq:kubo}
\end{equation}

\noindent where $\beta=1/k_B T$, $\langle \; \cdot \;\rangle = \left\{{\rm tr}[(\;\cdot\;) \exp(-\beta H)]\right\}/{\rm tr} [\exp(-\beta H)] $ denotes the thermodynamic expectation value at temperature $T$, $\Omega$ is the system's volume, and the currents are $J_a=\langle {\hat J}_a \rangle$, with $\hat{J}_a$ being the total current operator. Note that in extended systems, the operator ${\hat J}_a=\int_\Omega d\vec{r} {\hat j}_a(\vec{r})$ is an \emph{extensive} quantity, where ${\hat j}_a(\vec{r})$ is the current density operator, satisfying the continuity equation 

\begin{equation}
\frac{d\hat{\rho}_a(\vec{r},t)}{d t} = \frac{i}{\hbar}\,[H,\hat{\rho}_a] = 
-\nabla \cdot {\hat{j}}_a(\vec{r},t),
\label{eq:continuity}
\end{equation}

\noindent where $\hat{\rho}_a$ is the density of the corresponding conserved quantity, that is, electric charge for the electric current and energy for the energy current \footnote{The heat current is the difference between the total \emph{energy current} $J_u$ and the electrochemical potential energy current $\mu J_\rho$: $J_h=J_u-\mu J_\rho=J_u-(\mu/e)J_e$ \cite{callen}.}
and $\hbar$ is the reduced Planck's constant. Equation~(\ref{eq:continuity}) can be equally well written in classical mechanics, provided the commutator is substituted by the Poisson
bracket multiplied by the factor $i\hbar$. It can be shown that the real part of $L_{ab}(\omega)$ can be decomposed into a $\delta$-function at zero frequency defining a generalized \emph{Drude weight} $D_{ab}$ (for $a=b$ this is the conventional Drude weight) and a regular part $L_{ab}^{\rm reg}(\omega)$: 

\begin{equation}
{\rm Re} L_{ab}(\omega)=
2\pi D_{ab}\delta(\omega)+L_{ab}^{\rm reg}(\omega).
\label{eq:Lreg}
\end{equation}

\noindent The matrix of Drude weights can be 
also expressed in terms of time-averaged current-current correlations directly: 

\begin{equation}
D_{ab}=\lim_{\bar{t}\to\infty}\frac{1}{{\bar{t}}}\int_0^{\bar{t}} 
dt \lim_{\Omega\to\infty}\frac{1}{\Omega}\int_0^\beta d\tau\langle \hat{J}_a (0) \hat{J}_b (t+i\tau)\rangle.
\label{drudematrix}
\end{equation}

\noindent Note that it has been shown that non-zero  Drude weights,  ${D}_{ab}\ne  0$, are a signature of ballistic transport~\cite{Zotos1997,Zotos2004,Garst2001,H-M2005}, namely in the
thermodynamic limit the kinetic  coefficients $L_{ab}$ diverge linearly with the system size. 

The way in which the dynamic correlation functions in Eq.~(\ref{eq:kubo}) decay, determines the ballistic, anomalous, or diffusive character of the heat and charge transport, and it has been
understood that this decay is directly related to the existence of conserved dynamical quantities \cite{Zotos1997,Zotos2004}. For quantum spin chains and under suitable conditions,
it has been  proved that systems possessing conservation laws exhibit ballistic transport at finite temperature \cite{Ilievski2013}.

The following argument \cite{Benenti2013} highlights the role that conserved quantities play in the thermoelectric efficiency. The  decay of time  correlations for the currents can be related  to the existence of conserved  quantities by using \emph{Suzuki's formula} \cite{Suzuki1971}, which generalizes an inequality proposed by Mazur \cite{Mazur1969}. Consider a system of size $\Lambda$ along the direction of the currents (we denote its volume as $\Omega(\Lambda)$, and in the thermodynamic limit $\Omega\to\infty$) and Hamiltonian $H$, with a set of $M$ \emph{relevant conserved  quantities} ${Q}_m$, $m=1,\ldots  ,M$, namely the commutators  $[{H},Q_m]=0$. A constant of motion $Q_m$ is by definition relevant if it is not orthogonal to the currents under consideration, in our case $\langle \hat{J}_e Q_m \rangle \ne 0$ and $\langle \hat{J}_u Q_m \rangle \ne 0$.
It is assumed that the $M$ constants of motion are orthogonal, i.e., $\langle Q_m  Q_n\rangle = \langle Q_n^2 \rangle \delta_{mn}$ (this is always possible via a Gram-Schmid procedure).
Furthermore, we assume that the set $\{Q_m\}$ exhausts all relevant extensive conserved quantities. Then using Suzuki's formula, we can express the \emph{finite-size Drude weights}\footnote{Note that hereafter we shall use the simple thermal average correlator $\langle \hat{J}_a(0) \hat{J}_b(t) \rangle$ rather than the Kubo-Mori inner product $\int_0^\beta d\tau\langle \hat{J}_a (0) \hat{J}_b (t+i\tau)\rangle$; see Ref.~\cite{Ilievski2013} for a discussion of the assumptions needed to justify the use of the simple thermal-averaged expression.} 

\begin{equation}
d_{ab}(\Lambda)\equiv  \frac{1}{2\Omega(\Lambda)}
\lim_{\bar{t}\to\infty}\frac{1}{\bar{t}}
\int_0^{\bar{t}} dt \langle \hat{J}_a(0) \hat{J}_b(t) \rangle
\end{equation} 

\noindent in terms of the relevant conserved quantities: 

\begin{equation}
d_{ab}(\Lambda)=\frac{1}{2\Omega(\Lambda)}
\sum_{m=1}^M \frac{\langle \hat{J}_aQ_m \rangle \langle \hat{J}_bQ_m
  \rangle}{\langle Q_m^2 \rangle}.
\label{eq:finitesizedrude}
\end{equation}

\noindent On the other hand, the thermodynamic Drude weights can also be expressed in terms of time-averaged current-current correlations as

\begin{equation}\label{eq:Drude}
{D}_{ab}=\lim_{\bar{t}\to\infty}\lim_{\Lambda\to \infty}
\frac{1}{2\Omega({\Lambda}) \bar{t}}
\int_0^{\bar{t}} dt \langle \hat{J}_a(0) \hat{J}_b(t) \rangle .
\end{equation}

\noindent If the thermodynamic limit $\Lambda\to\infty$ commutes with the long-time limit $\bar{t}\to\infty$, then the thermodynamic Drude weights ${D}_{ab}$ can be obtained as 

\begin{equation}
{D}_{ab}=\lim_{\Lambda\to\infty} d_{ab}(\Lambda)\ .
\label{eq:drudeinfty}
\end{equation}

\noindent Moreover, if the limit does not vanish we can conclude that the presence of relevant conservation laws yields non-zero generalized Drude weights, which in turn imply that transport is ballistic, $L_{ab}\sim \Lambda$. As a consequence, the electrical conductivity is ballistic, $\sigma\sim L_{ee} \sim \Lambda$, while the thermopower is asymptotically size-independent,
$S\sim L_{eh}/L_{ee}\sim \Lambda^0$.

We can see from Suzuki's formula that for systems with a single relevant constant of motion ($M=1$), the ballistic contribution to $\det {\bm L}$ vanishes, since it is proportional to ${D}_{ee}{D}_{hh}-{D}_{eh}^2$, which is zero from Eqs. (\ref{eq:finitesizedrude}) and (\ref{eq:drudeinfty}). Hence, $\det {\bm L}$ grows slower than $\Lambda^2$, and therefore the thermal conductivity $\kappa\sim \det{{\bm L}}/L_{ee}$ grows sub-ballistically, $\kappa\sim \Lambda^\alpha$, with $\alpha<1$. Since $\sigma\sim\Lambda$ and $S\sim\Lambda^0$, we can conclude that $ZT\sim \Lambda^{1-\alpha}$ \cite{Benenti2013}. Hence $ZT$ diverges in the thermodynamic limit $\Lambda\to\infty$. This general theoretical argument applies for instance to systems where momentum is the
only relevant conserved quantity. Note that these conclusions for the thermal conductance and the figure of merit do not hold when $M>1$, as it is typical for completely integrable systems. In that case we have, in general, $D_{ee}D_{hh}-D_{eh}^2\ne 0$, so that thermal conductance is ballistic and therefore $ZT$ is size-independent. 

\begin{figure}[h!]
\center
  \centerline{\includegraphics[scale=0.45]{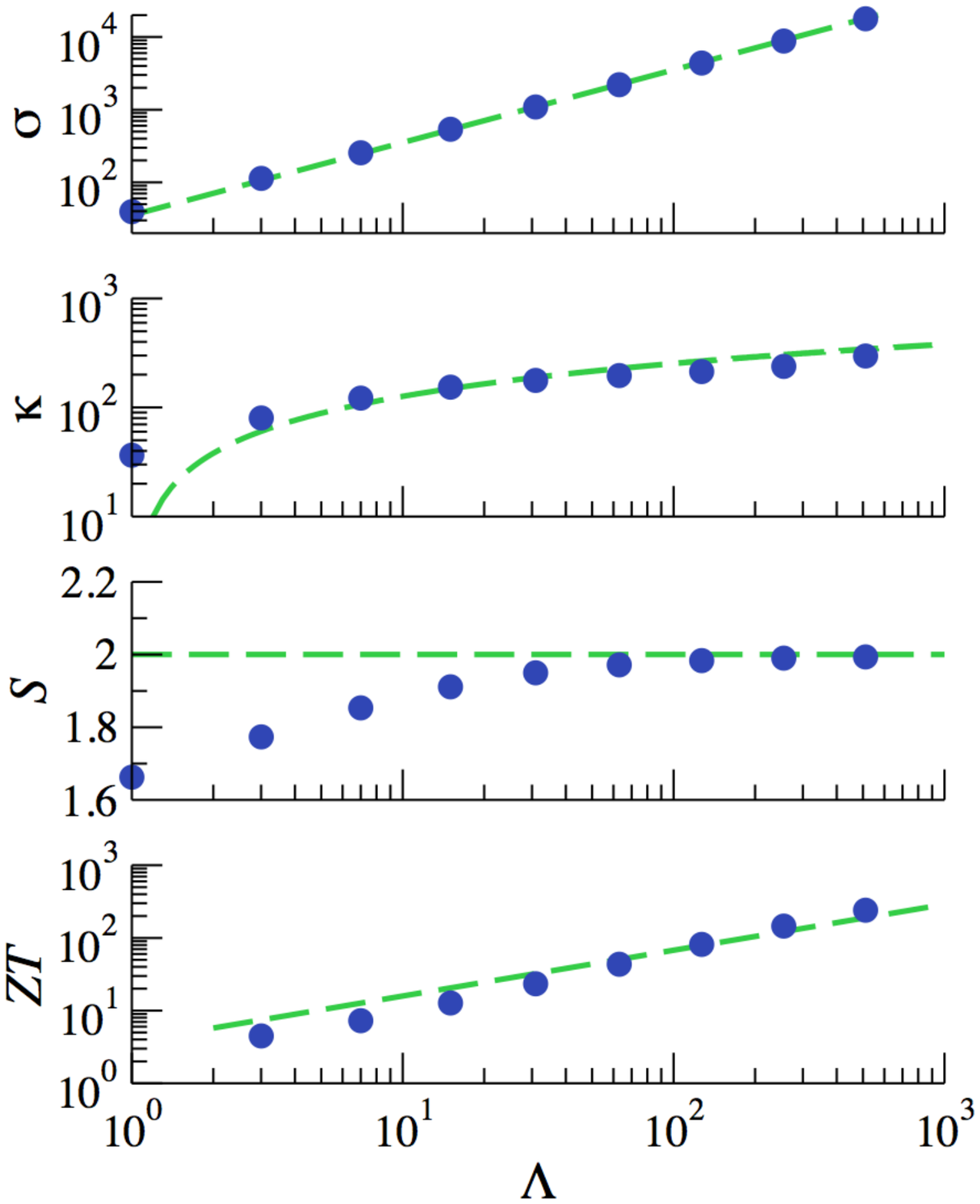}}
  \caption{Thermoelectric transport coefficients for the
    two-dimensional multiparticle collision dynamics gas
    of interacting particles as a function of
    the system size $\Lambda$
(for details see \cite{Benenti2014}). The dashed curves correspond from
top to bottom to $\sigma\propto \Lambda$,
$\kappa\sim \log\Lambda$, $S=2$, and $ZT\sim\Lambda/\log\Lambda$.}
  \label{fig:kapral}
\end{figure}

The above reasoning is not limited to quantum systems and has no dimensional restrictions; it has been illustrated by means of a diatomic chain of hard-point colliding particles \cite{Benenti2013}, where the divergence of the figure of merit with the system size cannot be explained in terms of the energy filtering mechanism \cite{Saito2010}, in a two-dimensional system connected to reservoirs \cite{Benenti2014}, with the dynamics simulated by the multiparticle collision dynamics method \cite{Kapral1999} and in a one-dimensional gas of particles with nearest-neighbor Coulomb interaction, modeling a screened Coulomb interaction between electrons \cite{Chen2015}. In all these (classical) models collisions are elastic and the component of momentum along the direction of the charge and heat flows is the only relevant constant of motion. Results for the two-dimensional multiparticle collision dynamics model are reported in Fig.~\ref{fig:kapral}. While the electrical conductivity grows ballistically, $\sigma\propto \Lambda$, the thermopower saturates to a value $S=2$ that can be also predicted analytically for this model \cite{Benenti2014}. Finally, the thermal conductivity grows according to the prediction of hydrodynamic theories \cite{lepri,dhar}, namely $\kappa\propto \log \Lambda$ in two dimensions. These results for the transport coefficients imply that the figure of merit diverges with the system size, $ZT\propto \Lambda/\log \Lambda$, and therefore the Carnot efficiency is achieved in the thermodynamic limit. 

We point out that it is a priori not excluded that there exist models where the long-time limit and the thermodynamical limit do not commute when computing the Drude weights. However, numerical evidence shows that for the models so far considered these two limits commute \cite{Benenti2013,Benenti2014,Chen2015}. Finally, we note that divergence of $ZT$ has been also predicted, on different theoretical considerations, for an ideal homogeneous quantum wire with weak electron-electron interactions, in the limit of infinite wire length \cite{matveev}.

\section{Discussion and concluding remarks}

Research in thermoelectricity remains widely regarded as a strategic activity in view of the critical problems related to energy production and storage, and considering that thermoelectric devices may be designed for specific purposes involving powers over a range spanning ten orders of magnitude: typically from microwatts to several kilowatts. However, despite all the money and efforts invested in the field of thermoelectricity over several decades, no one has managed yet to make decisive progress to break the glass ceiling over performance, thus enabling the much sought-after wide-scale applications. Non-interacting model systems provide a wealth of results and a solid socle to understand many aspects of the basic mechanisms that govern thermoelectric transport and energy conversion, and we saw that they also shed light on the reasons why in terms of performance of actual devices, we are still in the range of what became standard 30 years ago. 

It must also be said that the performance of a thermoelectric system does not rely solely on the intrinsic properties of the thermoelectric working fluid but also entails its interaction with its environment: poor thermal contacts with heat source and sink deteriorates significantly the overall energy conversion process. It is thus worthwhile to consider the optimization of the working conditions of thermoelectric devices, which necessitates a sound understanding of the coupling of these heat engines to their environment \cite{Apertet2012Conv,Apertet2012EPL,Apertet2014JAP} to ensure the highest possible efficiency at maximum output power. Finite-time thermodynamics is very well suited for such a purpose. Interestingly, thermoelectricity provides model systems that, in turn, are extremely useful in the development of theories in irreversible thermodynamics \cite{Groot1958}, and more generally in finite-time thermodynamics, which to date continues to attract much attention \cite{Andresen2011,Ouerdane2015b}. Indeed a number of outstanding questions in these fields related to finite-time optimization, may be advantageously tackled using thermoelectric systems as case studies since their basic operation allows for a physically transparent description of the phenomena at stake \cite{Gordon1991,Apertet2012PRE1,Apertet2013PRE3,Apertet2014PRE4}. 

However, optimization procedures for device operation, as necessary as they are, do not improve the fundamental energy conversion performed by the working fluid. The understanding of general mechanisms to improve thermoelectric efficiency by means of strongly interacting systems is only beginning to emerge. In particular, regimes near electronic phase transitions might be favorable for higher-efficiency thermoelectric conversion. Nonlinear momentum-conserving systems are also interesting, due to the ballistic nature of electrical conductivity combined with the anomalous behavior of thermal conductivity. Such dependence of thermal conductivity, growing slower than ballistically with the system size, is characteristic of the hydrodynamic regime. In this respect, it might be useful to remark that the hydrodynamic regime has been observed in graphene up to almost room temperature \cite{Polini}. 

To conclude, it is our hope that our theoretical results and analysis will help identify areas where genuine and significant progress is yet to be made to stimulate experimental research in the proposed directions.





\end{document}